\documentstyle[prd,aps,eqsecnum,epsf,psfig]{revtex}

\def\lesssim{\mathrel{\hbox{\rlap{\hbox{\lower4pt\hbox{$\sim$}}}\hbox{$<$}}}}
\def\gtrsim{\mathrel{\hbox{\rlap{\hbox{\lower4pt\hbox{$\sim$}}}\hbox{$>$}}}}
\def\alt{\mathrel{\hbox{\rlap{\hbox{\lower4pt\hbox{$\sim$}}}\hbox{$<$}}}}
\def\agt{\mathrel{\hbox{\rlap{\hbox{\lower4pt\hbox{$\sim$}}}\hbox{$>$}}}}

\def\mnras{MNRAS}
\def\apj{ApJ}
\def\prd{Phys.~Rev.~D}

\def\nat{Nature}

\newcommand {\be}       {\begin{equation}}
\newcommand {\ee}       {\end{equation}}

\begin{document}

\tighten

\title{Gravitational Waves from Neutron Stars with Large Toroidal B-Fields}
\author{Curt Cutler}
\address{Max-Planck-Institut fuer Gravitationsphysik\\
Am Muehlenberg 1, D-14476 Golm, Germany; cutler@aei.mpg.de}

\maketitle
\begin{abstract}
We show that NS's with large toroidal B-fields
tend naturally to evolve into potent gravitational-wave (gw) emitters. 
The toroidal field $B_t$ tends to distort the NS into a prolate shape, and  
this magnetic distortion dominates over 
the oblateness ``frozen into'' the
NS crust for $B_t \gtrsim 3.4\times 10^{12}{\rm G} ({\nu_s}/300\, {\rm Hz})^2$.
An {\it elastic} NS with frozen-in B-field of this magnitude
is clearly secularly unstable:
the wobble angle between the NS's angular momentum $J^i$ and the 
star's magnetic axis $n_B^i$ grow on a dissipation timescale until  
$J^i$ and  $n_B^i$ are orthogonal.
This final orientation is clearly the optimal one 
for gw emission. 
The basic cause of the instability is quite general, so 
we conjecture that the same final state is reached for a realistic NS, with
superfluid core. Assuming this,
we show that for LMXB's with
$B_t \sim 2 \times 10^{12}-2 \times 10^{14}$G, 
the spindown from gw's is sufficient to balance
the accretion torque---supporting a suggestion by Bildsten.
The spindown rates of most millisecond pulsars can also be
attributed to gw emission sourced by toroidal B-fields, 
and both these sources could
be observed by LIGO II. While the first-year spindown of a newborn 
NS is most likely dominated by electromagnetic processes, reasonable values of
$B_t$ and the (external) dipolar field $B_d$ can lead
to detectable levels of gw emission, for a newborn NS in our
own galaxy.
\end{abstract}
\twocolumn

%

\section{Introduction}\label{sec:intro}

Nearly periodic gravitational waves (gw's) from distorted or wobbling
neutron stars (NS's) are among the most promising sources for the
large-scale gravitational-wave detectors now coming online.
In this paper we show that NS's with large toroidal B-fields
tend naturally to evolve into potent gw emitters. 

While most known, young pulsars have dipolar magnetic field strengths
$B_d \sim 10^{12}\,$G, 
studies of soft-gamma repeaters and anomalous X-ray pulsars suggest
that a substantial fraction of NS's are born with larger magnetic fields, 
$B \sim 10^{14}-10^{15}$G~\cite{Lyutikov,Thompson,Stella,Chevalier}.
In cases where the B-field is initially ``wound up'' by 
differential rotation (as opposed
to convective motions), one would expect the
field's toroidal piece 
$B_t$ to dominate the poloidal part $B_p$.

The toroidal field $B_t$ tends to distort a NS into a prolate shape, and  
we show that this magnetic distortion dominates over 
the oblateness ``frozen into'' the
NS crust for $B_t \gtrsim 3.4\times 10^{12} ({\nu_s}/300 {\rm Hz})^2$G, 
where $\nu_s$ is the NS spin frequency.
An {\it elastic} NS with frozen-in B-field of this magnitude
is clearly secularly unstable:
the wobble angle between the NS's angular momentum $J^i$ and the 
star's magnetic axis $n_B^i$ grow on a viscous timescale until  
$J^i$ and  $n_B^i$ are orthogonal.
This final orientation is the optimal one 
for gw emission. 
The basic cause of the instability seems so general that 
we conjecture that the same final state is reached for a 
realistic NS (with crust and superfluid core).
In this paper we consider NS's where $B_t > B_p$, we assume that
most of the results for elastic NS's can be carried over
to the realistic NS case, and examine the consequences. 
We show that for LMXB's with
$B_t \sim 2 \times 10^{12}-2 \times 10^{14}$G, 
the spindown from gw's is sufficient to balance
the accretion torque--supporting a suggestion by Bildsten~\cite{Bildsten}.
The spindown rates of most millisecond pulsars can also be
attributed to gw emission sourced by toroidal B-fields 
of comparable size, $B_t \sim 10^{12}-10^{13}$G.
For several known millisecond pulsars, the gravitational waves 
would be detectable by
LIGO II. Finally, while the first-year spindown of a newborn 
NS is most likely dominated by electromagnetic (em) processes, 
reasonable values of $B_t$ and the (external) dipolar field $B_d$ can 
still lead to detectable levels of gw emission, for a newborn NS in our
own galaxy.

The idea that a large, toroidal magnetic field should drive the
magnetic axis orthogonal to $J^i$ is not new; a literature
search revealed that P. B. Jones had this idea $28$ years ago~\cite{jones1}. 
Jones~\cite{jones2,jones3,jones4,jones5}
went on to develop a scenario in which the dipole axes of young
pulsars (which he assumed have $B_t > B_p$) are {\it first} driven
orthogonal to $J^i$ by dissipative forces, on a timescale of $\sim
10^3$ yr, but then are slowly driven back parallel to $J^i$ by em
radiation reaction, for times $\agt 10^4$ yr. That is, Jones estimated
that at around age $10^3 - 10^4$ yr, dissipative forces would diminish
to the point that em backreaction would drive the dynamics.  Clearly,
Jones was strongly guided by the observation that most young
pulsars do {\it not} appear to have dipole axes orthogonal to $J^i$
(nor can they be perfectly aligned with $J^i$, as one would expect for
$B_p > B_t$, since they pulse). Given his strong argument that
dissipation should drive the magnetic axis orthogonal to $J^i$, and
the apparent contradiction by observation, there are three likely ways
out: 1) assume dissipative effects become negligible on a timescale of
order the ages of known, young pulsars, 2) assume the pulsar emission
region is not aligned along 
a principal axis of the {\it internal} B-field, or 3)
assume that the internal B-field is either not born with, or quickly
loses, any strong directionality.  Jones developed a model around the
first option, but found little support for it in the data.

Here we essentially resurrect the Jones argument, but concentrate on
a different class of objects possessing much smaller 
external dipole fields $B_d$:
the LMXB's and millisecond pulsars.
For these objects, it seems clear that dissipation must dominate
over the em torque in determining the alignment of the NS's axes.
We are agnostic on the question of why the {\it normal} (i.e., young, slow)
pulsars are generally neither aligned nor orthogonal rotators.
It is irrelevant to our argument whether option 1 or 2 above is correct;
option 3 seems the least likely to us, but if option 3 is always correct 
it would render this paper irrelevant.

Of course, it has been noted before that magnetically distorted NS's
can generate gw's; see, e.g., 
Bonazzola \& Gourgouhlon~\cite{Bonazzola_Gourgouhlon_96} and 
D.~I. Jones~\cite{jones00} 
and references therein. Also, Melatos \& Phinney~\cite{melatos_phinney} 
recently showed how the poloidal B-field on accreting NS's 
might build up ``mountains'' of accreted matter on the 
magnetic poles, leading to gw emission.
What is new here is our observation that 
NS's with large toroidal B-fields should
naturally evolve to configurations with large gw emission, and
our indentification of LMXB's and millisecond pulsars as
likely ``sites'' for this phenomenon.

In \S II we
derive the basic energetics and timescales. These are applied to 
LMXB's and millisecond pulsars in \S III, and to newborn NS's in \S IV. 
For our estimates, we shall always use the following
fiducial NS parameters: $M= 1.4 M_\odot$, $R = 10\,$km, and
$I = 10^{45}\,$g-cm$^2$.

\section{Energetics and Timescales}
Let $n_B^i$ be the NS's magnetic axis (so that if we call $n_B^i$ 
the $z-$axis, then the toroidal field $B_t$ points in the $\phi-$direction).
Following Pines \& Shaham~\cite{ps72}, we can write the  NS's inertia tensor
$I^{ij}$ as the sum of four pieces--a spherical piece and three quadrupolar
distortions--as follows:

\be\label{Iij}
I^{ij} = I_0\biggl[ 
e^{ij} + \epsilon_\Omega\, n_{\Omega}^in_{\Omega}^j
+ \epsilon_{d}\, n_d^i n_d^j
+ \epsilon_{B}\, n_B^i n_B^j\biggr]
\ee

\noindent Here $e^{ij}$ is the flat, spatial 3-metric, $I_0 e^{ij}$ is the
spherically symmmetric part of the inertia tensor, and
the terms proportional to $\epsilon_\Omega$, $\epsilon_d$, and 
$\epsilon_B$ are the quadrupolar distortions due to the star's spin,
crustal shear stresses, and the magnetic field, respectively. 

The centrifugal piece 
``follows'' the instantaneous (unit-)spin vector
$n_{\Omega}^i$, but the unit-vectors $n_d^i$ and $n_B^i$ 
are assumed fixed in the body frame. 
(However on long timescales, $n_d^i$ can be 
``re-set'' by crustal relaxation.)

The centrifugal piece $\epsilon_\Omega$ in Eq.~(\ref{Iij}) is approximately
given by $\epsilon_\Omega \approx 0.3 (\nu_s/{\rm kHz})^2$, where $\nu_s$ is the
spin frequency. The term  $\epsilon_d$ is due to the  
elastic crust's ``memory'' of some preferred shape. Absent the magnetic field, it is the residual oblateness the NS {\it would} have if it were spun down to 
zero frequency, without the crust breaking or otherwise relaxing.
(To understand this, consider a rotating NS with relaxed crust.
Removing the centrifugal force would 
decrease the NS's oblateness, but 
then shear stresses would also build up that tended to push the crust
back to its relaxed, oblate shape.) 
For a NS with relaxed crust, 
$\epsilon_d$ is proportional to the centrifugal oblateness: 
$\epsilon_d = b \epsilon_\Omega$. The coefficient $b$ was 
recently calculated by Cutler et al.~\cite{ucl},
who solved the coupled hydroelastic equations describing the
deformed crust and found $b \approx 2 \times 10^{-7}$.
(An older, back-of-the-envelope estimate of $b \sim 10^{-5}$ 
used by many authors turned out to be too high by
a factor $\sim 40$.) 
Therefore we adopt the estimate
\be\label{epsilon_d}
\epsilon_d \approx 6 \times 10^{-8} \biggl(\frac{\nu_s}{{\rm kHz}}\biggr)^2 \, .
\ee
Again, this estimate assumes the crust is nearly relaxed.
While this might not be a good assumption for young, slow NS's, 
it seems safe for the fast, old NS's of interest
to us here, for two reasons. First, during the LMXB phase, 
the entire crust is replaced by accreted matter as the NS spins up, and
it seems unlikely the new crust would ``remember'' the preferred shape 
from an earlier era. 
Second, the alternative requires the crust to withstand
average strains of order 
$3\times 10^{-3}(\nu_s/300\,{\rm Hz})^2$ 
for $\sim 10^7-10^{10}\,$yr 
without relaxing~\cite{ucl}, which
also seems unlikely.

Next we need $\epsilon_B$. Of course, the actual 
distortion produced by some toroidal field $B_t$
depends on the precise distribution of B-field (and hence currents) 
within the star, which is highly uncertain.
So we must be satisfied with
a rough estimate of $\epsilon_B$, which we derive as follows. 
Let $n_B^i$ be along the $z-$axis.
For an incompressible, constant-$\rho$ star that is not
superconducting, $\epsilon_B \equiv (I_{zz} - I_{xx})/I_{zz}$ 
depends only on the total
energy in the toroidal field~\cite{Wentzel,Ostriker_Gunn}:
\be\label{eb0} 
\epsilon_B = -\frac{15}{4} E_G^{-1} \int{\frac{1}{8\pi}B_t^2 \,dV} \, ,
\ee
\noindent where $E_G = (3/5)G M^2/R$ is the star's gravitational
binding energy. The sign of $\epsilon_B$ is negative 
because the toroidal magnetic field 
lines are like a rubber belt, pulling
in the star's waist at the magnetic equator.

Now, it is generally believed that for 
interior temperatures $T < 10^{9}-10^{10}\,{^\circ K}$, the protons in the NS interior are a type II superconductor; hence
the magnetic field is confined to flux tubes with field 
strength $B_{c1} \approx 10^{15}$G.
(For $B > B_{c1}$, the magnetic fluxoids overlap.)
Then for $B < B_{c1}$, the anisotropic part of the 
mean magnetic stress tensor is 
increased over its non-superconducting value  
by a factor $B_{c1}/B$, and $\epsilon_B$ increases by the
same factor~\cite{jones1,Easson_Pethick}.
(The virial-theorem-based derivation of 
$\epsilon_B$ in Ostriker \& Gunn~\cite{Ostriker_Gunn} 
makes this completely clear.) Putting this factor together
with Eq.~(\ref{eb0}), but using $E_G = (3/4)G M^2/R$ (i.e., the 
$n=1$ polytrope result, which quite accurately
gives the binding energy of realistic NS's), we adopt the estimate

\FL
\be\label{epsilon_B}
\epsilon_{B} = \left\{ \begin{array}{ll} 
-1.6 \times 10^{-6} (<B_t>/10^{15}{\rm G}) &\mbox{ $\ \ B_t < B_{c1}$,}\nonumber \\
-1.6 \times 10^{-6} (<B_t^2>/10^{30}{\rm G}) &\mbox{ $\ \ B_t > B_{c1}$,} \\
                \end{array}
        \right. 
\ee
\noindent where $<...>$ means ``volume-averaged over the NS interior.''
Of course, the first line of Eq.~(\ref{epsilon_B}) only applies
to NS's older than $\sim 0.1-1\,$yr (i.e., old and cold enough to be 
superconducting).

For simplicity we are assuming, here and below, 
that $B_t$ is sufficiently greater
than $B_p$ that we can neglect the latter. 
When is this a good approximation?
For an incompressible, constant-$\rho$, non-superconducting
NS, with $B^i$ of the form $B_p \hat z^i + B_t \hat\phi^i$ in the interior
and matched to a dipolar field in the exterior, the deformation 
is~\cite{Wentzel,Ostriker_Gunn}~\footnote{Ostriker \& Gunn~\cite{Ostriker_Gunn} 
actually give 3, not
2.1, as the coefficient of $<B_p^2>$ in Eq.~(\ref{pol_tor}), but
we claim they made an algebra error; the coefficient on the rhs of their
Eq.~(A19) should be $-7/40$, not their $-1/4$. With this correction, their
coefficent $3$ becomes $2.1$, in agreement with Wentzel~\cite{Wentzel}.}
\be\label{pol_tor}
\epsilon_B = \frac{25 R^4}{24 G M^2}\biggl(\frac{21}{10}<B_p^2> - <B_t^2>\biggr)\, .
\ee
For a superconducting NS, with $B_t, B_p << B_{c1}$, we claim this becomes
\be\label{pol_tor}
\epsilon_B = \frac{25 R^4}{24 G M^2}\biggl(2<B_p B_{c1}> - <B_t B_{c1}>\biggr) \, .
\ee
(The small change $(21/10) \rightarrow 2$ reflects the fact that the
{\it external} poloidal magnetic field energy becomes a negligible
fraction of the total field energy in the superconducting case.)
Clearly then, as long as $<B_t> \  \agt 4<B_p>$,
our estimates should be
fairly reliable.

We next consider the kinetic energy KE of the spinning NS as the
orientation of the body changes, for fixed angular momentum $J^i$.
While in the inertial frame the NS's angular velocity $\Omega^i$ clearly must 
remain approximately fixed (near the star's angular momentum $J^i$), 
$\Omega^i$ will migrate with respect to the body axes as the 
star precesses (and as the precession angle secularly changes).
Using $J^i = I^{ij}\Omega_j$, $KE = \frac{1}{2} J^i\Omega_i$, 
and expanding everything to lowest order in $\epsilon_\Omega$,
$\epsilon_d$ and $\epsilon_B$, we find:
\be\label{energy}
{\rm KE} = \frac{J^2}{2I_0}\biggl[1 - \epsilon_{\Omega} - 
\epsilon_d (n_d^i n_{J i})^2 -  \epsilon_B (n_B^i n_{J i})^2 \biggr]\, ,
\ee
\noindent where $n_J^i$ is the unit vector along $J^i$. 

Now consider the case where $\epsilon_B < 0 $ 
and $|\epsilon_B| >> \epsilon_d$, and
let $\theta$ be the ``wobble angle'' between $J^i$ and $n_B^i$.
Then the $\theta$-dependent piece of KE is $(2I_0)^{-1} J^2 (-\epsilon_B) {\rm cos}^2\theta$, which is clearly minimized for $\theta = \pi/2$.
(More precisely--including the effect of non-zero $\epsilon_d$-- 
KE is minimized for $|n_B^i n_{J i}| = {1\over 2}|(\epsilon_d/\epsilon_B) {\rm sin}2\beta|$, where $\beta$ is the angle between $n_B^i$ and $n_d^i$.)
In addition to the kinetic energy, the wobbling NS stores potential
energy (PE) in the stressed crust and the distorted fluid.  One easily
convinces oneself that the $\theta$-dependent pieces of PE and KE have the
same sign and same functional form ($\propto {\rm cos}^2\theta$), 
and that they have comparable magnitudes for NS's with relaxed crusts.
Therefore, considering
the total energy E = KE + PE leads to the same
conclusion about energy-minimization/stability 
as considering KE alone.

It is worth emphasizing that the NS need not actually 
be prolate to be unstable; indeed, for the cases of interest here
($\nu_s > 100\,$Hz), the centrifugal oblateness will be orders
of magnitude larger than $|\epsilon_B|$.
But what matters, for both precessional dynamics and stability, 
is the net prolateness/oblateness that is ``frozen into'' the star's
body frame, not the centrifugal piece 
that ``follows'' the star's angular velocity. 
Therefore the condition for
instability is $-\epsilon_{B} > \epsilon_d$, or 
\be\label{unstable}
B_t \agt 3.4 \times 10^{12}{\rm G} \biggr(\frac{\nu_s}{300\, {\rm Hz}}\biggr)^2 \, .
\ee

Assuming inequality (\ref{unstable}) is satisfied, how does the instability
grow? For an elastic star with frozen-in magnetic field, 
this is completely clear.
In the body frame, $\Omega^i$ precesses around $n_B^i$ 
with precession frequency $\nu_s \epsilon_B$ (neglecting 
a small correction due to nonzero $\epsilon_d$). 
Due to dissipation, the angle between
$\Omega^i$ and $n_B^i$ steadily increases until they are nearly orthogonal.
(For a review of NS precessional dynamics, 
see Cutler \& Jones~\cite{Cutler_Jones} and references therein.)

For a realistic NS, with superfluid core, the precessional dynamics could 
be more complicated. But as long as those dynamics do not 
reduce $-\epsilon_B$ below $\epsilon_d$ (i.e, by destroying the large-scale
coherence of the internal field or building up an equally large internal
poloidal piece), it seems inevitable, by simple energetics,
that dissipation will tend to drive $n_B^i$ orthogonal to $J^i$.
In what follows, we shall assume this is true. Some support for this
assumption comes from the work of Goldreich \& Reisenegger~\cite{Goldreich_Reisenegger}, who show that the composition of a NS interior is stably
stratified, impeding convective motions that might otherwise distort the 
interior B-field.

We next turn to timescales. There are four relevant ones: $\tau_{GW}$,
$\tau_{EM}$, $\tau_{ACC}$ and $\tau_{DIS}$. 
Once $n_B^i$ has been re-aligned perpendicular
to $J^i$, the NS spins down due to gw emission on a timescale 
$\tau_{GW} \equiv J/(2\dot J_{GW})$, given by 
\be\label{eq:tau_gw}
1/\tau_{GW} = 5.50 \times 10^{-13}{\rm s}^{-1} 
\biggl(\frac{\epsilon_B}{10^{-7}}\biggr)^2 \biggl(\frac{ \nu_s}{{\rm kHz}}\biggr)^4\, .
\ee

We must compare this to the timescale $\tau_{EM} \equiv J/(2\dot J_{EM})$ 
on which the NS spins down due to electromagnetic processes. 
Define $B_d \equiv 2{\cal M}/R^3$, where ${\cal M}$ is the NS's magnetic
dipole moment. (With this definition, $B_d$ is the value of the field
at the magnetic pole, on the NS surface, for an external field
that is perfectly dipolar.)   
Then  $\tau_{EM}$ is given by
\be \label{eq:tau_em}
1/\tau_{EM} = 4.88 \times 10^{-16}{\rm s}^{-1} 
\biggl(\frac{ B_d}{10^9{\rm G}}\biggr)^2 
\biggl(\frac{ \nu_s}{{\rm kHz}}\biggr)^2 \, .
\ee
For a magnetic dipole rotating in vacuum, the rhs of 
Eq.~(\ref{eq:tau_em}) would be multiplied by a factor 
${\rm sin}^2\alpha$, where $\alpha$ is
the angle between the spin vector and magnetic dipole direction. 
But Goldreich \& Julian~\cite{Goldreich_Julian} showed that a realistic NS is 
surrounded by plasma, and as a result Eq.~(\ref{eq:tau_em}) is a reasonable
estimate of the spindown rate even for an aligned (i.e., $\alpha =0$) rotator.

From Eqs.~(\ref{eq:tau_gw})-(\ref{eq:tau_em}), spindown due to gw 
emission dominates em spindown for
\begin{mathletters}
\begin{eqnarray}\label{gw_em}
B_t/B_d &>& 6.2 \times 10^3 \biggl(\frac{300 {\rm Hz}}{\nu_s}\biggr) \ \ , \ \   B_t < B_{c1}  \\ 
B_t/B_d &>& 2.5 \times 10^2 \biggl(\frac{300 {\rm Hz}}{\nu_s}\biggr)^{1/2}
\biggl(\frac{10^{14}{\rm G}}{B_d}\biggr)^{1/2},\   B_t > B_{c1}  
\end{eqnarray}
\end{mathletters}
Such a large ratio $B_t/B_d$ would not be expected in a normal pulsar,
but could occur in a LMXB/recycled pulsar.  For these sources, 
accretion apparently ``buries'' or otherwise reduces the 
external dipolar field, but the 
toroidal field in the deep interior could be left largely intact.

The em torque from the external dipole field 
causes the wobble angle $\theta$ between $J^i$ and
$n_B^i$ to damp (or grow, for ${\rm sin}^2\xi > 2/3$) 
on roughly the em spindown timescale $\tau_{EM}$~\cite{Goldreich}:
\be\label{align}
\frac{1}{{\rm sin}\theta}\frac{d\,{\rm sin}\theta}{dt} 
= -\frac{1}{2}{\rm cos}^2\theta (1 - \frac{3}{2}{\rm sin}^2\xi)/\tau_{EM} \, ,
\ee
\noindent where $\xi$ is the angle between the external dipole direction
and the magnetic axis $n_B^i$ of the internal B-field.
Similarly, 
gravitational radiation reaction tends to align the 
distortion axis $n_B^i$ with the spin direction (independent of the
sign of $\epsilon_B$) on the gw spindown timescale $\tau_{GW}$~\cite{Cutler_Jones}: $\dot \theta = -\frac{1}{32}\theta/\tau_{GW}$ for small 
wobble angle $\theta$, 
where here $\tau_{GW}$ is defined as the rhs of Eq.~(\ref{eq:tau_gw}).

Next there is the timescale $\tau_{ACC}$ on which accretion can
significantly change the NS's angular momentum (either magnitude or
direction). 
We approximate $\tau_{ACC} \equiv  J/2\dot J_{ACC}$ by~\cite{ucb00}
\be\label{tau_ACC}
1/\tau_{ACC} = 9.3 \times 10^{-16} {\rm s}^{-1} \biggl( \frac{\dot M}{10^{-9} M_\odot/{\rm yr}}\biggr) \biggl(\frac{300\, {\rm Hz}}{\nu_s}\biggr) \, .
\ee

Lastly, we consider the dissipation timescale $\tau_{DIS}$
on which the instability acts. Define $n$ to be the 
dissipation timescale $\tau_{DIS}$ divided by the wobble period, so  
$\tau_{DIS} \equiv n P/\epsilon_{B}$, where 
$\tau_{DIS} \sim |\epsilon_B|[J^2\theta^2/(2I)]/\dot E_{DIS}$.
We can rewrite this as
\be\label{eq:tau_dis}
1/\tau_{DIS} = 3.0 \times 10^{-8} {\rm s}\biggl(\frac{10^4}{n}\biggr)
\biggl(\frac{\nu_s}{300\,{\rm Hz}}\biggr)\biggl(\frac{\epsilon_B}{10^{-7}}\biggr).
\ee
\noindent
The $n$ factor is hard to estimate, but fortunately most of our
conclusions are essentially independent of $n$ over a huge range.
Previous authors who have considered precessional damping in NS's have 
generally concluded that precession damps quickly.
E.g., Chau \& Henriksen~\cite{ch71} 
considered dissipation in the elastic
crust as $\Omega^i$ (and hence the centrifugal bulge) precess
through the body frame. 
The crust is periodically distorted by the wobbling motion, and
the elastic energy stored in the crust is
some fraction $F$ of the total wobble energy, where we
estimate $0.05 < F < 0.5$. (For the Earth, F is $0.11$~\cite{Stacey}.)
If some fraction $1/Q$ of the elastic energy is dissipated in each
wobble period, then $n \sim Q/F$. Typically $Q \sim 10^4$ for 
terrestrial metals, so one might estimate $n \alt 10^5$.
Other mechanisms seem likely to lead to faster damping:
the core and crust will not wobble together rigidly, and 
relative internal motions will lead to frictional damping, e.g., at
the crust-core interface, or through interactions of neutron superfluid
vortices with electrons and nuclei in the inner crust.
(E.g., Alpar \& Sauls~\cite{as88} estimated $n \approx 10^2- 10^4$ for slowly
rotating NS's, with the dissipation due to electron scattering 
off pinned superfluid vortices. However superfluid vortices 
certainly will {\it not} remain pinned in the wobble regimes of
interest to us--large wobble angle and $\nu_s \agt 300$ Hz--
so this particular estimate does not seem immediately relevant to our 
case; see Link \& Cutler 2002~\cite{Link_Cutler}.)
We therefore leave $n$ as unknown parameter to be determined by observation 
or improved theory. The point is that, for LMXB's and millisecond pulsars, 
any $n$ less than $\sim 10^{9}$  
means the dissipation timescale is much 
shorter than all the others, and so should 
successfully drive $n_B^i$ orthogonal to $J^i$.

We have collected all the necessary formulae. We now apply them to
interesting cases: LMXB's, millisecond pulsars, and newborn NS's.

\section{LMXB's and millisecond pulsars}

\subsection{Evolution Scenario}

The arguments and timescales above lead to the following
evolution scenario for (some large fraction of) 
LMXB's and millisecond pulsars. 
The NS is born with $B_t \agt 4 B_p$ and 
$B_d \sim 10^{12}-10^{14}$G. At birth,  $n_B^i$ is likely
nearly aligned with the NS's spin (and so is nearly aligned with
$n_d^i$). The NS spins down
electromagnetically, but much later it is recycled
by accretion from a companion. Accretion
reduces the exterior field $B_d$ below $\sim 10^9$G;
however in the interior, $B_t$ remains in the
range $\sim 10^{12}-10^{15}$G. As the NS spins back up and
its external B-field decays, it reaches a state where 
$\tau_{DIS}$ is shorter than the other timescales
(including $\tau_{EM}$, thanks to the decay of $B_d$),
while $|\epsilon_B| > \epsilon_d$.
Then dissipation rapidly 
``flips''  $n_B^i$ perpendicular to $J^i$.
Initially $n_d^i$ flips along with $n_B^i$, but
we expect the crust of a rapidly spinning NS will
crack/relax such that its new preferred axis $n_d^i$ is again
aligned with $\Omega^i$, but now is perpendicular to $n_B^i$.
This final configuration is clearly a minimum of both
KE and PE (for fixed J), independent of the
relative sizes of $\epsilon_d$ and $|\epsilon_B|$.
Thus while the NS continues to spin up, dissipation
will ensure that $n_d^i$ remains aligned with $J^i$, with
$n_B^i$ perpendicular to both. This is important because
$\epsilon_d$ increases quadratically with spin frequency, and
so (as we shall see) eventually surpasses $|\epsilon_B|$ for 
the fastest LMXB's and millisecond pulsars. Our point is that as long
as $n_d^i$ has already been ``re-set'' along $J^i$ {\it before}
that happens, then $n_B^i$ will remain orthogonal to $J^i$.
After accretion stops, the NS continues to spin down by gw emission.
And because $n_B^i$ is time-varying (rotating around $J^i$ with
frequency $\nu_s$) while $n_d^i$ is not, 
it is the magnetic distortion that sources the gravitational waves.

A potential objection to this scenario arises if one assumes
that the pulsar beam is aligned with the axis $n_B^i$ of the
B-field in deep interior, since then we would predict
that (a large fraction of) millisecond pulsars should be 
orthogonal rotators, which
is not observed. So for our picture to be sensible, we should  
assume that the B-field near the NS surface evolves (e.g., via 
accretion and crust-cracking) in such a way that that the 
pulsar beam is {\it not} aligned with $n_B^i$.



\subsection{LMXB's}

The distribution of LMXB spin periods suggest
that many have ``hit a wall'' at $\nu_s \sim 260-590\,$Hz--far
below the likely maximum rotation rate of $1-1.5\,$kHz~\cite{Bildsten,ucb02}. 
Bildsten~\cite{Bildsten} proposed
that they had reached an equilibrium where accretion torque was
balanced by spindown from gravitational radiation (resuscitating an idea of Wagoner's~\cite{Wagoner}).

Equilibrium between accretion torque and gw spindown
implies an ellipticity 
\be\label{equil}
\epsilon = 4.5 \times 10^{-8} \biggl( \frac{\dot M}{10^{-9} M_\odot/{\rm yr}}\biggr)^{1/2} \biggl(\frac{300 {\rm \, Hz}}{\nu_s}\biggr)^{5/2} \, .
\ee
For the LMXB's with known spin rates, the range of $\dot M$'s 
is $\sim 10^{-11}-2\times 10^{-8}M_\odot/$yr, with 
implied $\epsilon$'s in the range
$\sim 3 \times 10^{-9}-3\times 10^{-7}$~\cite{ucb02}.

On this assumption, gw's from the brightest LMXB's
(especially Sco X-1) should be detectable by LIGO II.
Ushomirsky et al.~\cite{ucb00} showed that crustal ``mountains''
can provide the required ellipticity, but only if the 
crustal breaking strain
$\sigma_{max}$ is larger than $\sim 10^{-3}-10^{-2}$.  
Toroidal B-fields place
no such requirement on the crust's breaking strain:
$B_t \sim 2\times 10^{12}- 2\times 10^{14}$G 
provides the required ellipticity. Our picture requires that
gw torque dominates em torque, 
or $B_d \alt 5\times 10^{8}-5\times 10^{10}$G, by Eq.~(\ref{gw_em}).
There are no direct measurements of $B_d$ in LMXB's
to compare this to, but $B_d \alt 10^9\,$G is indicated by 
the lack of X-ray pulses. 

While for simplicity we have assumed $B_t >> B_p$, it should 
be clear that the basic picture remains the same if
they are comparable, so long as the total $\epsilon_B$ is negative 
and $|\epsilon_B| \sim 10^{-8}-10^{-7}$.



\subsection{Millisecond Pulsars}
We turn now to millisecond radio pulsars (which we define
as those with spin periods $P < 10$ ms), which
are commonly assumed to be the descendants of
LMXB's. Our picture suggests than many should now
be spinning down by gw emission. The NS's characteristic
age $\tau_c \equiv P/(2\dot P)$ is then just $\tau_{GW}$.
We can re-write Eq.~(\ref{eq:tau_gw}) as
\be\label{epsB}
|\epsilon_B| = 6.0\times 10^{-9}\biggl(\frac{10^{10}{\rm yr}}{\tau_c}\biggr)
\biggl(\frac{P}{5\, {\rm msec}}\biggr)^2 \, , 
\ee
\noindent or  $B_t \sim 4 \times 10^{12}$G
for our fiducial millisecond pulsar values ($\tau_c = 10^{10}\,$yr and
$P = 5\,$msec)--consistent with the lower end of the range we
inferred for LMXB's. (A comparison of derived $\epsilon_B$'s 
in the LMXB and millisecond pulsar populations would be interesting, but
is clearly complicated by selection effects; 
e.g., low-$|\epsilon_B|$ millisecond pulsars 
live longer, and so will dominate the observed sample.)  
Note that $|\epsilon_B| > \epsilon_d $ only
for $P \agt 4.0\, {\rm msec} (\tau_c/10^{10}{\rm yr})^{1/2}$ 
(from Eqs.~\ref{epsilon_d} and \ref{epsB}), so the oldest/fastest
millisecond pulsars would not be unstable in the way we described in \S II.
However, as explained in \S III.A, that probably does not matter:
their magnetic axes were flipped orthogonal to $J^i$ while they
were being spun up, and will not flip back if the crust
has relaxed.

Given a millisecond pulsar at distance D, spinning down
by gw emission, it is straightforward to compute the
detectability of the gw signal $h(t)$. 
The rms $S/N$ (averaged over source direction and polarization) is 
\be
S/N = 1.1 \biggl(\frac{1{\rm kpc}}{D}\biggr) 
\biggl(\frac{T_o}{1{\rm yr}}\biggr)^{1/2} \biggl(\frac{10^{10}{\rm yr}}{\tau_c}\biggr)^{1/2}
\biggl(\frac{2\times 10^{-24}}{S_h^{1/2}(f)}\biggr)  
\, ,
\ee
where $f=2\nu_s$ is the gravity-wave frequency, $S_h(f)$ is the (single-sided) 
noise spectral density, and $T_0$ is the observation time.
\noindent 
The broad-band LIGO-II noise curve has a broad minimum
at $f_{gw} \sim 400\,$ Hz, with $S_h^{1/2}(f) 
\approx 2\times 10^{-24}$  there~\cite{Thorne}.
Using this fiducial value for $S_h^{1/2}(f)$, ones finds there
are at least 
$4$ known millisecond pulsars with $S/N > 2.7$ in a 1-yr observation.
PSR J0437-4715 ($f_{gw} = 347\, Hz$, $\tau_c = 0.6 \times 10^{10}$yr, 
and $D = 0.18$ kpc~\cite{Toscano}) 
gives the highest S/N ratio: $S/N = 7.9\,$. 
Also, PSR J1744-1134 ($f_{gw} = 491\,$ Hz, 
$\tau_c = 0.72 \times 10^{10}$yr, $D = 0.36$ kpc~\cite{Toscano2}) 
yields $S/N = 3.6$; PSR J1024-0719 ($f_{gw} = 387\,$ Hz, 
$\tau_c = 2.7 \times 10^{10}$yr, $D = 0.20$ kpc~\cite{Toscano})
yields $S/N = 3.3\,$; and PSR J1012+5307 ($f_{gw} = 381\,$Hz, 
$\tau_c = 0.6 \times 10^{10}$yr, $D = 0.52$ kpc~\cite{Lorimer})
gives $S/N = 2.7$.
Factor $\sim 2.5$ improvements in these SNR's 
could be obtained by operating the detector in narrow-band, 
signal-recycling mode.

Note that since the gw signal is almost completely known
from radio observations (i.e., up to polarization, amplitude and 
overall phase), $S/N \agt 2.7$ is enough for confident detection.

\section{Newborn pulsars}

As another application of our basic idea, consider the spindown
gw's from a supernova in
our own galaxy ($D \sim 10$ kpc).
(The supernova rate in our galaxy is $\sim 1/40$yr,
so this author expects to be around for the next one.) 
It does not seem outlandish to posit field strengths 
$B_d \sim 10^{14}$G and 
$B_t \sim 10^{15}$G. 
Now em radiation dominates the spindown torque, but
there is significant gw emission as well. 

The matched-filtering $(S/N)^2$ for a NS spinning down {\it solely}
due $l=m=2$ gw's is~\cite{Owen_Lindblom}
\be\label{sn}
\biggl(S/N\biggr)^2 = \frac{2 G}{5 \pi D^2 c^3}\int_{f_{min}}^{f_{max}}\frac{dJ/df}{f\, S_h(f)}df \, .
\ee
For our case, we just need to 
multiply the numerator in Eq.~(\ref{sn}) by $\dot J_{GW}/\dot J_{EM} = 
\tau_{EM}/\tau_{GW} \propto f^2$. Approximating the high-frequency 
part of the broad-band, LIGO-II noise 
spectrum by $S_h(f) =  2.1 \times 10^{-47} {\rm s} (f/10^3{\rm Hz})^2 $~\cite{Thorne}, we find
\be
S/N = 11.7 \biggl(\frac{10 \,{\rm kpc}}{D}\biggr) \biggl(\frac{\epsilon_B}{10^{-6}}\biggr)\biggl(\frac{10^{14}{\rm G}}{B_d}\biggr) \bigl[{\rm ln}\bigl(\frac{f_{max}}{f_{min}}\bigr)\bigr]^{1/2}\, .
\ee
\noindent Here we can take $f_{min}$ to be the frequency ($\sim 500~$Hz) 
where the actual LIGO-II broad-band noise curve ``flattens out.''

E.g., consider a NS with 
$B_t = 2\times 10^{15}$G  and $B_d = 10^{14}$G,
which spins down from $\nu_s \sim 1$ kHz 
to $\nu_s = 250$ Hz in $\sim 10$ days
(and to $\nu_s \approx 80$ Hz after one year).
In order for $n_B^i$ to ``flip'' before the
NS spins down substantially, we need $\tau_{DIS} < \tau_{EM}$, or
$n \alt 1.3 \times 10^3$, 
by Eqs.~(\ref{eq:tau_em}) and (\ref{eq:tau_dis}).
Assuming $n$ {\it is} this small, and 
taking $[{\rm ln}(f_{max}/f_{min})]^{1/2} \approx 1$, we find
$S/N = 75$ for $D= 10\,$kpc, 
which would likely be detectable 
even in a semi-blind search--i.e., one where neutrino 
detectors and other channels give the collapse time and
a rough position, but where there is no radio signal to give us 
the rotational phase as function of time.

\acknowledgements
It is a pleasure to thank Lars Bildsten, Ian Jones,  
Bernard Schutz, Chris Thompson, and Greg Ushomirsky for helpful discussions.



\begin{references}

\bibitem{Lyutikov}
M. Lyutikov, C. Thompson, and S.~R. Kulkarni, in {\it Neutron Stars in Supernova Remnants} (ASP, 2002); astro-ph/0111319.

\bibitem{Thompson}
C. Thompson, astro-ph/0110679.

\bibitem{Stella}
S. Mereghetti, L. Chiarlone, G.~L. Israel, and L.Stella; astro-ph/0205122.

\bibitem{Chevalier}
R.A. Chevalier, in {\it Neutron Stars in Supernova Remnants},
(ASP, 2002); astro-ph/0201295.

\bibitem{Bildsten}
L. Bildsten, \apj \ {\bf 501}, L89 (1998).

\bibitem{jones1}
P.~B. Jones, Ap. and Sp. Sci. {\bf 33}, 215 (1975).

\bibitem{jones2}
P.~B. Jones, \nat \ {\bf 262}, 120 (1976).

\bibitem{jones3}
P.~B. Jones, \apj \ {\bf 209}, 602 (1976).

\bibitem{jones4}
P.~B. Jones, Ap. and Sp. Sci. {\bf 45}, 669 (1976).

\bibitem{jones5}
P.~B. Jones, \mnras \ {\bf 178}, 87 (1977).

\bibitem{Bonazzola_Gourgouhlon_96}
S. Bonazzola and E. Gourgoulhon, A\&A
{\bf 312}, 675 (1996).

\bibitem{jones00}
D.I. Jones, Ph.D. thesis, University of Wales, Cardiff, 2000.

\bibitem{melatos_phinney}
A. Melatos and E.~S. Phinney, PASA {\bf 18}, 421 (2001).

\bibitem{ps72}
D. Pines and J. Shaham,  Nature Physical Science {\bf 235}, 43 (1972); Phys. Earth Planet. Interiors {\bf 6}, 103 (1972)

\bibitem{ucl}
C. Cutler, G. Ushomirsky, and B. Link, in preparation.

\bibitem{Cutler_Jones}
C. Cutler and D.~I. Jones, \prd \ {\bf 63}, 24002 (2001).

\bibitem{Wentzel}
D.~G. Wentzel, ApJ Supp. {\bf 47}, 187 (1960).

\bibitem{Ostriker_Gunn}
J.~P. Ostriker and J.~E. Gunn, ApJ {\bf 157}, 1395 (1969). 

\bibitem{Easson_Pethick}
I. Easson and C.~J. Pethick, \prd \ {\bf 16}, 275 (1977).

\bibitem{Goldreich_Reisenegger}
P. Goldreich and A. Reisenegger, \apj \ {\bf 395}, 250 (1992).

\bibitem{Goldreich_Julian} 
P. Goldreich and W.H. Julian, \apj \ {\bf 157}, 869 
(1969).

\bibitem{Goldreich}
P. Goldreich, \apj \ {\bf 160}, L11 (1970).

\bibitem{ucb00} 
G. Ushomirsky, C. Cutler, and L. Bildsten, \mnras \ {\bf 319}, 902 (2000).

\bibitem{ch71}
W.Y. Chau and R.N. Henriksen, Astrophys. Letters {\bf 8}, 49 (1971)

\bibitem{Stacey}
F.D. Stacey, \emph{Physics of the Earth, 3rd Ed.} (Brookfield Press, 1992).

\bibitem{as88}
A. Alpar and J.A. Sauls, \apj \ {\bf 327}, 723 (1988)

\bibitem{Link_Cutler}
B. Link and C. Cutler, to appear in MNRAS; astro-ph/0108281.

\bibitem{ucb02} 
G. Ushomirsky, L. Bildsten, and C. Cutler, in \emph{Proceedings of the Third
Eduoardo Amaldi Conference} (AIP Press, 2000); astro-ph/0001129.

\bibitem{Wagoner}
R.~V. Wagoner, \apj \ {\bf 278}, 345 (1984).

\bibitem{Owen_Lindblom}
B.~J. Owen and L. Lindblom, Class. Quant. Grav. {\bf 19}, 1247 (2002).

\bibitem{Thorne}K.~S. Thorne, LIGO Internal Document G000025-00-M.

\bibitem{Toscano}
M. Toscano et al., \mnras \ {\bf 307}, 925 (1999); astro-ph/9811398.

\bibitem{Lorimer}
D. Lorimer, Living Reviews in Relativity (2001); astro-ph/0104388.

\bibitem{Toscano2}
M. Toscano et al., \apj \ {\bf 523}, L171 (1999); astro-ph/99906372.


\end{references}
\end{document}